\documentclass[12pt]{article}
\usepackage{graphics}
\usepackage{epsfig}
\usepackage{hyperref}
\usepackage{color}
\usepackage{graphics}
\usepackage{graphicx}
\usepackage{latexsym,amssymb}
\title{Competing jump cycles for vacancy diffusion in binary alloys}
\author{Zerihun Getahun$^1$, Mesfin Asfaw$^2$ and Mulugeta Bekele$^1$\\ $^1$Department of Physics, Addis Ababa University
 P.O.Box 1176,\\ Addis Ababa, Ethiopia\\ $^2$Department of Physics and  Graduate Institute of\\ Biophysics National Central University  Jhongli, 32054  Taiwan}

\begin{document}
\maketitle

\date{Received: date / Revised version: date}
\abstract{ The mean-first-passage-times (MFPTs) for a vacancy that diffuses (via one-
and six-jump cycles) in a two dimensional ordered binary alloy are evaluated using the
properties of random walks on networks.  We investigate the effect of temperature and relative barrier height on
the ratio between the MFPTs of the two cycles. At low temperature we find that the
six-jump cycle takes shorter time while at high temperature the one-jump cycle
takes shorter time than that of the six-jump cycle for the range of parameters considered.}

 %\PACS{vacancy diffusion, binary alloy, six-jump cycle  }

\section{Introduction}

The mechanism by which a single vacancy diffuses in mono-atomic crystalline material
is basically through site exchange with one of its nearest-neighbor atoms. The vacancy
diffusion continues through the material successively in a random way such that each
new site occupied is usually energetically identical to any other earlier occupied
site. As such, order is maintained throughout the vacancy diffusion.

The situation is not so simple if the crystalline material is, for instance, composed
of a binary alloy which consists of two interpenetrating simple cubic sublattices that
are predominantly occupied by two different atoms, A and B.
\begin{figure}
\epsfig{file=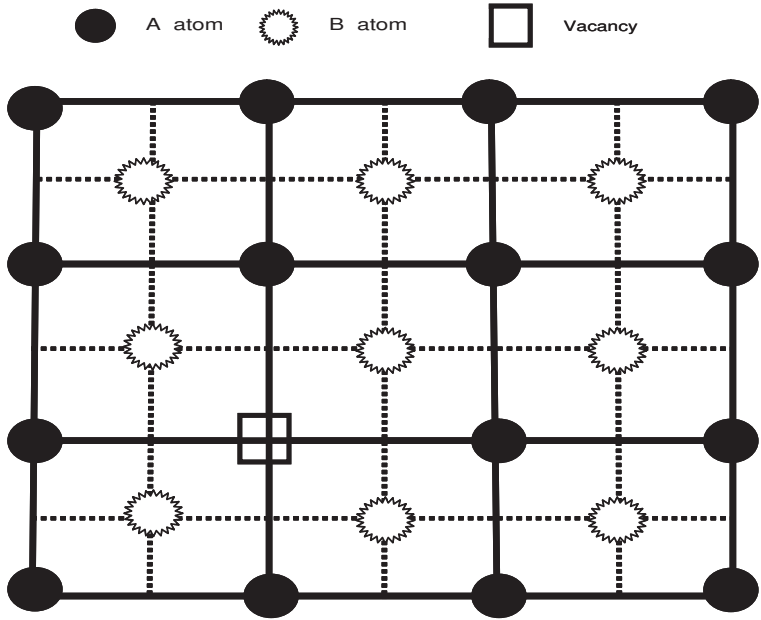,width=10cm} \caption{A binary alloy on two-dimensional lattice
with a single vacancy occupying a site on sublattice A.}
\end{figure}
 Figure 1 is an
illustration of a binary alloy in two-dimensional lattice with a single vacancy.
Whenever the vacancy on site A exchanges its site with one of its nearest-neighbor
atoms, the process leads to disorder in the crystalline structure. If the vacancy
randomly moves successively via nearest-neighbor jumps, a string of anti-structure
atoms would lead to disorder in the material. To avoid this problem of disordering,
two alternative diffusion mechanisms are likely to take place: either jumps of the
vacancy to further distant sites on the same sublattice or a cycle of successive
intermediate jumps to nearest-neighbor in which the atomic disorder appearing during
the earlier part of one cycle is followed by successive healing during the later part
of the cycle. The first alternative is usually called one-jump cycle. The prominent
cycle for the second alternative is the six-jump cycle. It was first suggested by H.
B. Huntington and later discussed by Elcock and McCombie \cite{el1}.
\begin{figure}
\epsfig{file=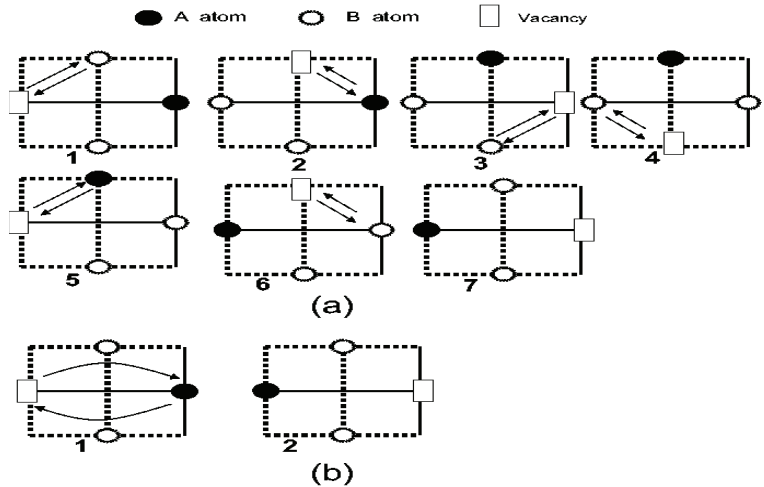,width=10cm} \caption{A possible path for (a) six-jump cycle and
(b) one-jump cycle.}
\end{figure}
Figures 2a and
2b illustrate the possible paths for six- and one-jump cycles, respectively, for a
binary alloy in a two-dimensional lattice.

A previous work dealt with calculating diffusion coefficient via six-jump cycles using rate equation method \cite{dra}. An even earlier work \cite{ar1}used mean-first passage method to compare the diffusion coefficients of the A and B atoms. In our present work, we raise a specific question regarding the times required for a vacancy to diffuse via the two dominant cycles and compare them using appropriate parameters.

Which one of the two alternatives will the vacancy prefer to diffuse through the
lattice? One way to answer this question is to compare the times taken for the vacancy
to go from one stable state to the next stable state via the two alternatives.

The problem of vacancy jump from one site to the next through site exchange with an
atom can be seen as a barrier crossing problem of an idealized Brownian particle
representing the combined vacancy-atom pair involved in the site exchange. The
direction of vacancy motion can be taken as the direction of movement of the Brownian
particle. Since each state to which the vacancy jumps is either stable or metastable,
it is reasonable to consider the next jump process as independent of its previous one.
In other words, we will consider the Brownian particle to have enough time to get
thermalized at each site before taking the next jump.

The problem of evaluating the times the vacancy takes from one stable site to the next
stable site via the one- and six-jump cycles depend on the corresponding potential
energy profiles of the two alternative paths. These times are usually called mean-
first-passage-times (MFPTs). In the case of the one-jump cycle, the vacancy has to
make a single jump over a high barrier to the nearest site on the same sublattice. In
the case of the six-jump cycle, the vacancy has to undergo through six local jumps
where the total barrier height is subdivided to ultimately arrive at a nearest site on
the same sublattice through a longer path than that of the one-jump cycle.
\begin{figure}
\epsfig{file=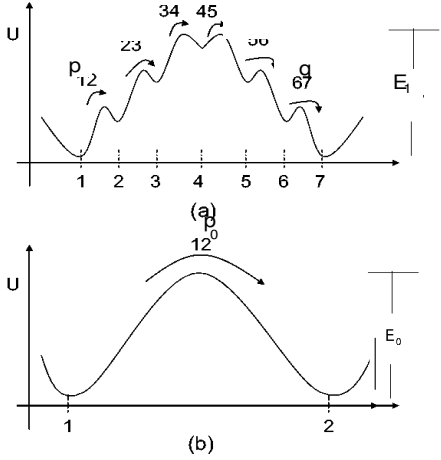,width=10cm} \caption{Schematic diagram showing energy profile for
(a) six-jump cycle and (b) one-jump cycle. Note local jump probabilities $p$ (up) and $q$ (down) in (a) and jump probability $p_0$ in (b)).}
\end{figure}
 Figs. 3a
and 3b are the energy profiles for the six  and one-jump cycles, respectively,
during the course of each cycle. Given the specific potential profile for the one-jump
cycle, one can use the standard method of solving Brownian diffusion in a potential
field to evaluate the MFPT for the one-jump cycle. On the other hand, to evaluate the
MFPT for the six-jump cycle we use a technique that has first been formulated by
Goldhirsch and Gefen \cite{gold1,gold2} and later applied by one of us in
collaboration with others  \cite{m1}. This technique requires knowledge of local jump
probabilities between successive sites as an input in order to evaluate its MFPT.

In this work we consider the vacancy diffusion  on a two-dimensional lattice. In this case,
there are four possible nearest stable sites on the same sublattice to which the
vacancy can jump by either performing the one- or the six-jump cycle. In principle,
the vacancy can attempt to diffuse via all these paths, select one of the paths and ultimately reach one of the
four sites in one cycle.

The rest of the paper is organized as follows. In section 2,
we  consider the multiple paths scenario for the two alternative jump cycles that describe
vacancy diffusion on two-dimensional lattice and determine their corresponding closed
form expressions for their MFPTs. Section 3 compares the values of MFPTs as a
function of some of the parameters of interest and explores the different
possibilities defining them. In section 4, we present the summary and conclusion.

\section{MFPTs along  the one- and six-jump cycles}

The MFPT,$\tau_1$, that  the Brownian particle takes to complete the one-jump cycle along a single path is worked out in the Appendix. We consider a potential  profile similar to  Fig. 3b but for simplicity the potential profile is  considered to be  a  piece-wise linear potential as in Fig. 6 of the Appendix. The expression for the MFPT in the low temperature regime
turns out to be (see Eq. (A10) in the Appendix)
\begin{equation}
\tau_1 = \frac{1}{D}\left(\frac{k_BTa}{E_0}\right)^2 e^{\frac{E_0}{k_BT}}
\end{equation}
where $a$  is the distance between nearest neighbors, $E_0$ is the barrier height energy, $D$ is the
diffusion coefficient, $k_B$ is Boltzman's constant and $T$ is the temperature of the crystal medium.
One should note that there are four possible stable lattice sites to which the vacancy jumps on the same sublattice.
If the vacancy chooses the one-jump cycle, the MFPT $\tau_{14}$ it takes to reach one of the four possible sites on the same sublattice is simply
\begin{equation}
\tau_{14}= \frac{\tau_1}{4}= \frac{1}{4D}\left(\frac{k_BTa}{E_0}\right)^2 e^{\frac{E_0}{k_BT}}.
\end{equation}
On the other hand, if the vacancy chooses the six-jump cycle to ultimately arrive at one of the four  stable lattice sites on the same sublattice there are two possible paths. Therefore, there are a total of eight possible paths  which the vacancy  can select to move by one sublattice distance. The evaluation of the MFPT, $\tau_{68}$, the vacancy takes to move by one sublattice distance via
the six-jump cycle after choosing one of the eight possible paths is done using the technique formulated by  Goldhirch
and Geffen \cite{gold1,gold2} and given by
\begin{equation}\label{eq:22}
       \tau_{68}=\frac{-q(p+q)}{4p}\left(\frac{2p^3}{q^5}+\frac{-7p^2-9pq+q^2}{q^2(p+q)^2}\right),
\end{equation}
where $p$ and $q$ are  the local jump  probabilities (up and down) over the rugged potential  corrseponding to the six-jump cycle (see $p$ and $q$ in Fig. 3a)
The closed form expressions  for $p$ and $q$ are derived in the Appendix.

                \section{Result and Discussion}

The expressions for the MFPTs are functions of a few physical quantities and include
the barrier heights of the one- and six- jump cycle, $E_0$ and $E_1$, the background
thermal energy, $k_BT$, of the binary alloy, the local barrier heights, $\epsilon_1$
and $\epsilon_2$, for the six-jump cycle (see Fig. 6 in Appendix), the diffusion coefficient $D$ and the
spacing, $a$, between the sublattices. In order to compare the MFPTs between the two
alternatives we first identify two dimensionless quantities that are parameters
controlling the MFPTs. One is the dimensionless quantity $b=\frac{\epsilon_1}{k_BT}$
that controls the local hopping rate for the six-jump cycle. The other dimensionless
quantity is the ratio $r=\frac{E_0}{E_1}$ which compares the total barrier
corresponding to the one-jump cycle with that of the total barrier height for the
six-jump cycle (see Figs. 3 (a) and (b)).

Let us define a dimensionless quantity, $f_m$, that compares the MFPT, $\tau_{14}$, for the one-jump
cycle to that of the MFPT, $\tau_{68}$ for the six-jump cycle given by
\begin{equation}
f_m = |\log(\frac{\tau_{14}}{\tau_{68}})|.
\end{equation}
Fig. 4  shows three plots of  $f_{m}$ as a function of $b$ (scaled local barrier height for the six-jump cycle)
corresponding to three different $r$ values that compare the total barrier heights of the two cycles:$r=0.8, 1 and 1.2$.
Each plot has a region of negative slope as well as a region of positive slope. Note that the region of negative slope
corresponds to a situation where the MFPT, $\tau_{14}$, via the one-jump cycle is {\it smaller} than that of the
MFPT, $\tau_{68}$, via the six-jump cycle. This implies that in the high temperature regime the one-jump cycle takes
shorter time than that taken by the six-jump cycle. This is because when the background thermal kick is high enough,
the vacancy can easily cross the barrier height through one-jump cycle. On the other hand, the region of positive slope
corresponds to a situation where $\tau_{68}$ is {\it smaller} than $\tau_{14}$. This implies that at low temperature
the six-jump cycle takes shorter time than that of the one-jump cycle. This is because, at low temperature, the
background thermal kick is weak for the vacancy to cross the
barrier via one-jump cycle. On the other hand, for the six-jump cycle there are six small barrier heights which can be
crossed with relatively small thermal kicks. Thus at low temperature the vacancy can cross the relatively small local
barriers quickly compared to the large barrier of the one-jump cycle.
At the inflection point of the plots, $f_m$ is zero and this corresponds to a situation where the MFPT via both cycles
is the {\it same}. This clearly shows that at a certain temperature the MFPT for a vacancy diffusing through one-jump
cycle is equal to that of vacancy diffusion via the six-jump cycle.
Comparing the three plots, when $r=E_0/E_1$ gets large the MFPT via the six-jump cycle predominantly gets relatively shorter
compared to that of the one-jump cycle.
\begin{figure}
\epsfig{file=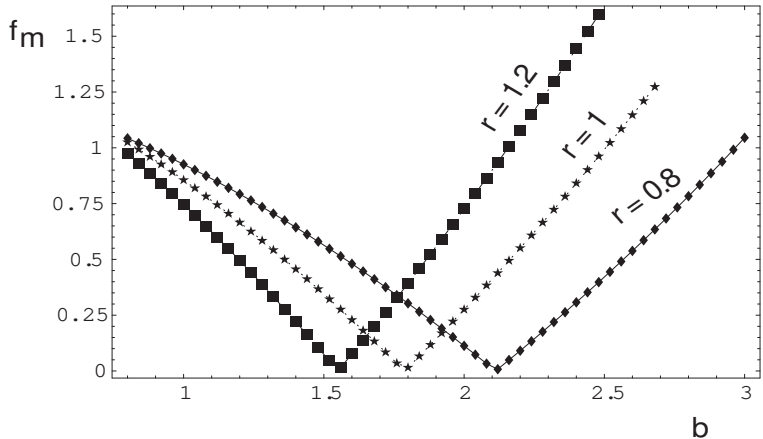,width=10cm}
\caption{ Plot which shows $f_{m}$ versus $b$ for fixed $r=0.8$, $r=1$ and $r=1.2$. }
\end{figure}
\begin{figure}
\epsfig{file=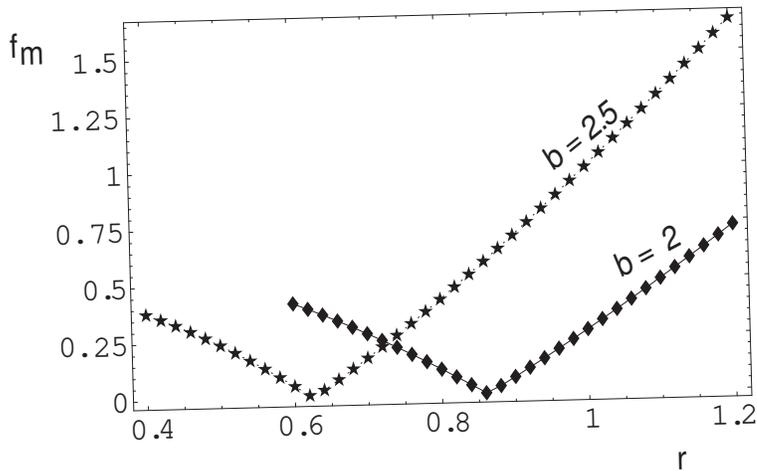,width=10cm}
\caption{Plot which demonstrates $f_{m}$ versus $r$ for fixed $b=2.5$ and $b=2$. }
\end{figure}
Fixing $b=2$ and $b=2.5$, we investigate how the $f_{m}$  behaves as a function of $r$ as shown in Fig. 5. 
The figure clearly demonstrates that the MFPT for one-jump cycle becomes shorter than the  MFPT  of six-jump cycle when  
the thermal background temperature is strong enough  as long as  the potential barrier for one-jump cycle, $E_0$, is 
smaller than $E_1$ for six-jump cycle.

         \section{Summary and conclusion}

      We studied the MFPT for a vacancy diffusing in two dimensional binary alloys. We investigated how the MFPT behaves as a function of the two model parameters. The central result we obtained shows that the
six-jump cycle takes invariably shorter time than that of the one-jump cycle for the parameter ranges considered at low temperature regimes. When the back ground temperature   is strong enough,  the one-jump cycle is dominant.

As a concluding remark, there are two issues we would like to point out for future consideration. The first one concerns the need to explore other ranges of parameters not considered in this work. The second one is the need to relate our findings to experimental results.

        \section*{\bf Acknowledgement}
Z.G and M.B. would like to thank The Intentional Program in Physical Sciences, Uppsala University, Uppsala, Sweden for the support they have been providing for our research group. M.A. would like to thank  Hsuan-Yi Chen for stimulating discussions.

\renewcommand{\theequation}{A\arabic{equation}}
  % redefine the command that creates the equation no.
  \setcounter{equation}{0}  % reset counter
\section*{Appendix}

            In this Appendix we derive the values for $p$, $q$ and $p_0$ in terms of the  model parameters. For a Brownian particle  that moves in a highly viscous medium under the influence of external potential
             $V(x)$, the Langevin
        equation that governs  the dynamics of such a particle is given
        by:
         \begin{equation}
        \frac{dx}{dt}= -\frac{V'(x)}{\gamma}  +
        \xi(t)\sqrt{\frac{2k_{B}T}{\gamma}}
        \end{equation}
    where $\gamma$ and $T$ denote the constant friction coefficient and the temperature, respectively. $V(x)$ is the external potential, $ k_{B}$ is the Boltzman's constant and $  \xi(t)$ is the delta correlated noise term.
The  corresponding  Fokker-Planck equation can be written as,
         \begin{equation}\label{eq:8}
        \partial_{t}p(x,t)=\partial_{x}[{V'(x)\over \gamma} p(x,t)] +
        D\partial^2_{x}p(x,t).
        \end{equation}
        Here $p(x,t)$  represents  the probability of a particle to be found at a position $x$ at
        time $t$, and $D= k_{B}T/\gamma$ is the diffusion coefficient. Starting from Eq. A2, one can derive the expression for MFPT in a bistable potential whose inverse is the jump probability \cite{ga}.
        Taking a piecewise linear potential which is described by
    \begin{figure}
        \epsfig{file=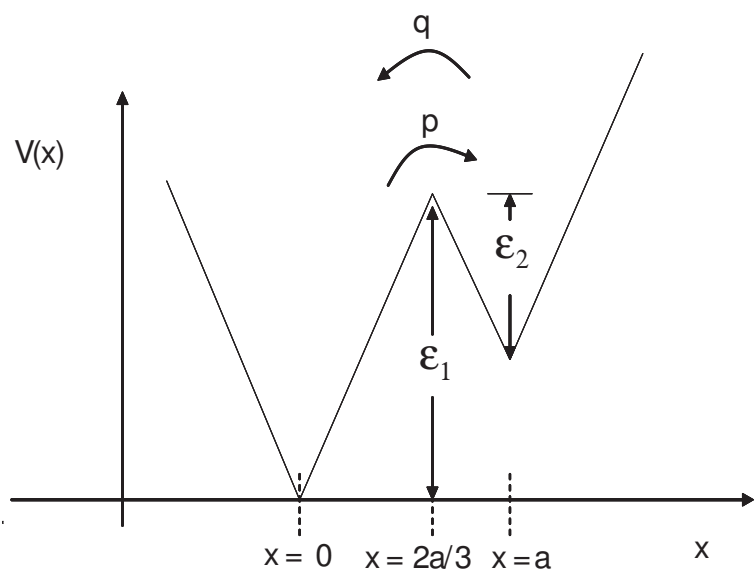,width=10cm}
         \caption{Asymmetric piecewise linear double well potential. }
         \end{figure}
   \begin{equation}
V(x)=\cases {\frac{-3\epsilon _{1}x}{2a},& \textrm{$ x\leq 0$};\cr
\frac{3\epsilon _{1}x}{2a},& \textrm{$0\leq x\leq {2a\over 3}$};\cr
\frac{-3\epsilon _{1}x}{2a}+2\epsilon _{1},& \textrm{${2a\over 3}\leq x\leq a$};\cr\frac{3\epsilon _{1}x}{2a}-\epsilon _{1},& \textrm{$x>a$},\cr}
\end{equation}
and noting that $\epsilon_2=\epsilon_1/2$, the MFPT $\tau(0\longrightarrow a)$ to reach $x=a$,  starting from $x=0$ is
\begin{equation}
         \tau (0\longrightarrow a) =({2a^{2}e^{{-\epsilon_{1}\over 2{k_BT}}} (k_BT)^2\over 9D \epsilon_{1}^2})(4-4 e^{{\epsilon_{1}\over {k_BT}}}+8 e^{{3\epsilon_{1}\over {2k_BT}}}- e^{{\epsilon_{1}\over {2k_BT}}}({\epsilon_{1}\over k_BT}+8 )).\end{equation}
When $\epsilon_{1}$ is large compared to $k_BT$,  the above expression takes a simple form:
\begin{equation}
         \tau (0\longrightarrow a)= \frac{4}{D}\left(\frac{k_BTa}{\epsilon_1+ \epsilon_2}\right)^2 e^{\frac{\epsilon_1}{k_BT}}.
\end{equation}
For the case $\epsilon_{1}$ is large compared to $k_BT$,  probability $p$  to jump from $x=0$ to $x=a$ is the inverse of $\tau(0\longrightarrow a)$ so that
       \begin{equation}
         p=\frac{D}{4}\left(\frac{\epsilon_1+ \epsilon_2}{k_BTa}\right)^2 e^{\frac{-\epsilon_1}{k_BT}}.
         \end{equation}
     Similarly for the reverse case, the MFPT  $\tau(a\longrightarrow 0)$ for the vacancy to reach $x=0$ from $x=a$ is given by
     \begin{equation}\label{eq:7}
         \tau(a\longrightarrow 0)=({2a^{2}e^{{-2\epsilon_{2}\over {k_BT}}} (k_BT)^2\over 9D \epsilon_{1}^2})(4-4 e^{{\epsilon_{2}\over {k_BT}}}+8 e^{{3\epsilon_{2}\over {k_BT}}}+ e^{{2\epsilon_{2}\over {k_BT}}}({2\epsilon_{2}\over {k_BT}}-8 )). \end{equation}
For low temprature regime the above equation takes a simple form:
\begin{equation}\label{eq:7}
         \tau(a\longrightarrow 0)= \frac{4}{D}\left(\frac{k_BTa}{\epsilon_1+ \epsilon_2}\right)^2 e^{\frac{\epsilon_2}{k_BT}}
         \end{equation}
 and the associated local jump probability $q$ is given by
\begin{equation}
         q=\frac{D}{4}\left(\frac{\epsilon_1+ \epsilon_2}{k_BTa}\right)^2 e^{\frac{-\epsilon_2}{k_BT}}.
         \end{equation}
If the bistable potential is symmetric with large barrier height $E_{0}$ (compared to thermal energy) and width $a$,
the MFPT  taken to jump in both directions is the same and is given by

                    \begin{equation}\label{eq:8}
          \tau_1 = t(0\longrightarrow a)= \frac{1}{D}\left(\frac{k_BTa}{E_{0}}\right)^2 e^{\frac{E_{0}}{k_BT}}
 \end{equation}
               while the corresponding jump  the probability, $p_{0}=q=p$ takes the value
                \begin{equation}\label{eq:8}
                p_{0}={D}\left(\frac{E_{0}}{k_BTa}\right)^2 e^{\frac{-E_{0}}{k_BT}}.
                \end{equation}

            \end{document}